\documentclass[12pt]{article}
\usepackage{amssymb}
\usepackage{amsmath}
\usepackage{amsfonts}
\usepackage{graphicx}
\usepackage{color}

\newcommand{\sign}{\textrm{sgn}}
\title{Nanomechanics of a screw dislocation in a functionally graded material using the theory of gradient elasticity}
\author{K.M. Davoudi$^a$\footnote{Corresponding author. Tel:  1-617-496-8145 Email: davoudi@seas.harvard.edu},  H.M. Davoudi$^b$, E.C. Aifantis$^{c,d}$,}

\begin{document}
\date{}
\maketitle

\begin{center}
\small{
$^a$ School of Engineering and Applied Sciences, Harvard University, Cambridge, MA 02138, USA\\
$^b$ Department of Mathematics, Tarbiat Moallem University, Tehran, Iran\\
$^c$ Center for the Mechanics of Material Instabilities, and Manufacturing Processes (MMIMP), Michigan Technological University,
Houghton, MI 49931, USA\\
$^d$ Laboratory of Mechanics and Materials (LMM), Polytechnic School, Aristotle University of Thessaloniki, P.O. Box 468,
Thessaloniki 54124, Greece\\
}
\end{center}

\abstract{The modest aim of this short article is to provide some new results for a screw dislocation in a functionally graded material within the theory of gradient elasticity. These results, based on a displacement formulation and the Fourier transform technique, extends earlier findings (Lazar, M., 2007. On a screw dislocation in a functionally graded material, Mech. Res. Comm. 34, 305-311) obtained by the stress function method, to the case of the second gradient elasticity. Rigorous and easy-to-use analytical expressions for the displacements, the strains and the stresses are obtained which are free from singularities at the dislocation line.}

\textbf{keywords: }{Screw dislocation; Functionally graded material; Second strain gradient elasticity}
\section{Introduction}
The stress function technique was employed by Lazar [1] to derive stress and strain fields for a screw dislocation in a functionally graded material by using the theory of first gradient elasticity. Analytical non-singular expressions were derived for the strains and the (first order) stresses, but the double stresses remained singular. Such results are derived here by using the Fourier transform technique which, in addition, provides exact analytical expressions for the displacement field. Moreover, the problem is reconsidered within the framework of ``second strain gradient elasticity" which eliminates the singularities from the double stress expressions as well. Recent work on gradient elasticity [2,3] has revealed the need of using higher order gradients of strain in the stress-strain relation in order ro interpret experimental results pertaining to dislocation density tensor and more accurately describe the details of the relevant stress/strain fields near the core of dislocations contained in small volumes. This is the case in particular, for dislocations contained in functionally graded materials (FGMs), the use of which has advocated since mid 80's in relation to ultra high temperature and ultra high weight requirements for aircraft, space vehicles and
other applications. Generally FGMs refer to heterogeneous composite materials, in which mechanical properties are intentionally made to vary smoothly and continuously from point to point. This is controlled by the variation of the volume fraction of the constituent materials. Ceramic/ceramic and metal/ceramic are typical examples of FGMs ([4,5] and references quoted therein). Although several aspects of FGMs have been reviewed comprehensively ([6-8] and references quoted therein) only few
investigations have been made to assess the role of the dislocations in FGMs.

With the exception of [1], the classical theory of linear elasticity was routinely utilized to calculate the elastic fields produced by defects (dislocations and disclinations) in FGMs. However, classical continuum theories are
scale invariant in which no intrinsic length appears and so fails
when one attempts to explain the nano-scale phenomena near
defects. As a result, elastic singularities are present in these solutions and size effects which dominate in small volumes cannot be captured. These undesirable features are removed within the second strain gradient elasticity formulation presented in this paper. The solutions obtained herein by using the Fourier transform technique and a displacement formulation are reduced to the corresponding expressions of classical elasticity and first gradient elasticity as, for example, were obtained in [1] through the use of the stress function approach. As a result, analytical expressions for the displacement field (in addition to those for the stresses and strains) are derived. The extra dividend is the derivation of non-singular expressions for the double stresses which diverge near the dislocation core in the first order strain gradient theory.

\section{Classical Solution }
We consider a screw dislocation with Burgers vector  $\textbf{b}=(0,0,b_z)$ in an infinite medium with a varying shear modulus $\mu=\mu(y)=\mu_0 e^{2ay}$ ($a\geq0$) in the framework of classical elasticity. This is a problem of anti-plane shear with the only non-vanishing component of displacement $u_z^0(x,y)$ satisfying the displacement equilibrium equation
\begin{equation}
\left(\nabla^2+2a\frac{\partial}{\partial y}\right)u_z^0=0,
\end{equation}
where $\nabla^2$ denotes the Laplacian. Using the substitution $u_z^0=w^0\,e^{-ay}$, we obtain
\begin{equation}
\left(\nabla^2-a^2\right)w^0=0,
\end{equation}
which by means of the Fourier transform
$$\tilde{f}(s)=\mathfrak{F}\{f(x);x\rightarrow
s\}=\int_{-\infty}^{\infty}f(x) e^{-\textrm{i}sx}dx,$$
where $\textrm{i}=\sqrt{-1}$, is reduced to the following ordinary differential
equation
\begin{equation}
\left(-s^2-a^2+\frac{d^2}{d y^2}\right)\widetilde{w}=0.
\end{equation}
Since $u_z^0$ is finite everywhere, we arrive
at
\begin{align}
\tilde{u}_z^0=e^{-ay}\left\{ {\begin{array}{*{20}c}
   {A(s)e^{ - y\sqrt {s^2  + a^2 } } } & {(y > 0)},  \\
   {B(s)\,e^{y\sqrt {s^2  + a^2 } } } & {(y < 0)},  \\
\end{array}} \right.
\end{align}
where  the two unknown functions, $A(s)$ and $B(s)$, are constants with respect to $y$. In view of the present dislocation configuration, we have
\begin{align}
u_z^0(x,0^+)-u_z^0(x,0^-)&=b_z H(-x),\\
\notag \varepsilon_{zy}^0(x,0^+)=\varepsilon_{zy}^0(x,0^-)&\Leftrightarrow
\frac{\partial u_z^0}{\partial y}(x,0^+)=\frac{\partial
u_z^0}{\partial y}(x,0^-),
\end{align}
where $\varepsilon_{zy}^0$ is the classical strain, and $H(-x)$ is the Heaviside step function. Taking the Fourier transform of the above conditions, we have
\begin{align}
\tilde{u}_z^0(x,0^+)-\tilde{u}_z^0(x,0^-)=b_z\left(\pi\delta(s)+\frac{\textrm{i}}{s}\right);\quad
\frac{\partial \tilde{u}_z^0}{\partial y}(s,0^+)=\frac{\partial
\tilde{u}_z^0}{\partial y}(s,0^-),
\end{align}
and, thus, the unknown functions $A(s)$ and $B(s)$ are determined as
\begin{align}
A(s)=\frac{-\textrm{i}a}{2s\sqrt{s^2+a^2}}+\frac{\textrm{i}}{2s},\quad
B(s)=\frac{-\textrm{i}a}{2s\sqrt{s^2+a^2}}-\frac{\textrm{i}}{2s}-\pi\,\delta(s).
\end{align}
It follows that
\begin{equation}
\tilde{u}_z^0={b_z}\,e^{-ay}\left[\frac{-\textrm{i}a}{2s\sqrt{s^2+a^2}}
e^{-|y|\sqrt{s^2+a^2}}+\sign(y)\,\frac{\textrm{i}}{2s}\,e^{-|y|\sqrt{s^2+a^2}}-\pi\,\delta(s)\,H(-y)\,e^{ay}\right].
\end{equation}
and by taking the inverse Fourier transform, i.e.
$$u_z^0(x,y)=\mathfrak{F}^{-1}\{\tilde{u}_z^0(s,y);s\rightarrow
x\}=\frac{1}{2\pi}\int_{-\infty}^{\infty}\tilde{u}_z^0(s,y)\,e^{\textrm{i}sx}ds,$$
we obtain (in view of the symmetry of the integral) the final expression
\begin{align}
u_z^0=\frac{b_z}{2\pi}\,e^{-ay}&\int_0^\infty
\frac{a\sin{(sx)}}{s\sqrt{s^2+a^2}}\,e^{-|y|\sqrt{s^2+a^2}}ds\\
\notag &-\sign(y)\,\frac{b_z}{2\pi}\,e^{-ay}\int_0^\infty
\frac{\sin{(sx)}}{s}\,e^{-|y|\sqrt{s^2+a^2}}ds-\frac{b_z}{2}\,H(-y).
\end{align}
Next we note that in small strain theory the (compatible) total strain $\varepsilon_{ij}^{T}$ may be written as
$$\varepsilon_{ij}^{T}=(1/2)(u_{i,j}+u_{j,i})=\varepsilon_{ij}+\varepsilon_{ij}^{P},$$
where $\varepsilon_{ij}$ and $\varepsilon_{ij}^{P}$ denote the usual (incompatible) elastic and plastic strains, respectively. It follows that
\begin{align}
&\varepsilon_{zx}^{0T}=\frac{1}{2}\,\frac{\partial u_z^0}{\partial
x}=\frac{b_z}{4\pi}\,e^{-ay}\int_0^\infty
\frac{a\,\cos{(sx)}}{\sqrt{s^2+a^2}}\,e^{-|y|\sqrt{s^2+a^2}}ds\\
\notag
&\hspace{2.7cm}-\sign(y)\,\frac{b_z}{4\pi}\,e^{-ay}\int_0^\infty
\cos{(sx)}\,e^{-|y|\sqrt{s^2+a^2}}ds,\\
\notag &\varepsilon_{zy}^{0T}=\frac{1}{2}\,\frac{\partial u_z^0}{\partial y}
=\frac{b_z}{4\pi}\,e^{-ay}\,\int_0^\infty \frac{s
\sin{(sx)}}{\sqrt{s^2+a^2}}\,e^{-|y|\sqrt{s^2+a^2}}ds\\
\notag &\hspace{2.7cm}-\frac{b_z}{2\pi}\delta(y)\int_0^\infty
\frac{\sin(sx)}{s}ds+\frac{b_z}{4}\,\delta(y).
\end{align}
With the help of the identities
$$\int_0^\infty \frac{\sin (sx)}{s}ds=\frac{\pi}{2}\sign(x),\quad
\int_0^\infty \frac{\cos
(sx)}{\sqrt{s^2+a^2}}\,e^{-|y|\sqrt{s^2+a^2}}ds=K_0(ar),$$
where $r=\sqrt{x^2+y^2}$ and $K_n$ denotes the modified Bessel function of the second kind and of order $n$, the integrals
appearing in Eq.~(10) can readily be evaluated to give
\begin{align}
\varepsilon_{zx}^{0T}=\frac{b_z}{4\pi}\,e^{-ay}\,\left[a\,K_0(a
r)-\frac{a
y}{r}\,K_1(a r)\right],\quad
\varepsilon_{zy}^{0T}=\frac{b_z}{4\pi}\,e^{-ay}\,\frac{a x}{r}
K_1(a r)+\frac{b_z}{2}\delta(y)\,H(-x).
\end{align}
The last term in Eq.~(11)$_2$ which is singular on the half-plane $y=0$ and $x\leq 0$, corresponds to the plastic strain $\varepsilon_{zy}^{0P}=b_z\delta(y)H(-x)/2$. The other term on the right hand side of Eq.~(11)$_2$ may thus be regarded as the elastic
strain. Using the constitutive law, $\sigma_{zi}^0=2\mu\varepsilon_{zi}^0$ ($i=x,y$), the stresses read
\begin{align}
\sigma_{zx}^0=\frac{b_z\,\mu_0}{2\pi}\,e^{ay}\,\left[a\,K_0(a
r)-\frac{a
y}{r}\,K_1(a r)\right],\quad
\sigma_{zy}^0=\frac{b_z\,\mu_0}{2\pi}\,e^{ay}\,\frac{a x}{r} K_1(a
r).
\end{align}
which are the same as those earlier obtained in [1] by the stress function approach, and which are singular at the dislocation line.

\section{Strain gradient elasticity solution}
Within a simplified theory of linearized anisotropic theory of second strain gradient elasticity proposed in [9] (for a corresponding form of first strain gradient elasticity and a robust method for solutions of corresponding boundary value problems, the reader may consult [10,11]), the strain energy density has the form
\begin{align}
W=\frac{1}{2}\,C_{ijkl} \varepsilon_{ij} \varepsilon_{kl}+\frac{1}{2}\,\ell^2 C_{ijmn} \varepsilon_{mn,k} \varepsilon_{ij,k}+\frac{1}{2}\,\ell'^4 C_{ijmn} \varepsilon_{mn,kl} \varepsilon_{ij,kl},
\end{align}
where $\varepsilon_{ij}$ is the elastic strain tensor, $\ell$ and $\ell'$ are internal lengths, and $C_{ijkl}$ is the stiffness  tensor of the form
$$C_{ijkl}=\lambda(\boldsymbol{x})\delta_{ij}\delta_{kl}+\mu(\boldsymbol{x})\left(\delta_{ik}\delta_{jl}+\delta_{jk}\delta_{il}\right),$$
with the Lam\'{e} constants $\lambda(\boldsymbol{x})$ and $\mu(\boldsymbol{x})$ being given functions of the spatial coordinates. The corresponding expressions for the elastic-like first order stress ($\sigma_{ij}^E$) and the higher-order double ($\tau_{ijk}$) and triple ($\tau_{ijkl}$) stresses are given by
$$\sigma_{ij}^E:=\frac{\partial W}{\partial \varepsilon_{ij}}=C_{ijkl}\varepsilon_{kl},\
\tau_{ijk}:=\frac{\partial W}{\partial \varepsilon_{ij,k}}=\ell^2\, C_{ijmn}\varepsilon_{mn,k},\
\tau_{ijkl}:=\frac{\partial W}{\partial \varepsilon_{ij,kl}}=\ell'^4\,C_{ijmn}\varepsilon_{mn,kl}.$$
while the Cauchy stress $\sigma_{ij}$ (note that in the notation of [1] this was denoted by $\sigma^0_{ij}$ and termed total stress) satisfies, in the absence of body forces, the usual equilibrium equation
\begin{equation}
\sigma_{ij,j}=\sigma_{ij,j}^E-\tau_{ijk,kj}+\tau_{ijkl,klj}=0.
\end{equation}

For the present case of anti-plane shear we have
$$\sigma_{zj}^E=2\,\mu\epsilon_{zj},\quad
\tau_{zjk}=2\,\ell^2\,\mu\epsilon_{zj,k},\quad
\tau_{zjkl}=2\,\ell'^4\,\mu\epsilon_{zj,kl}; \hspace{1cm}(j,k,l=x,y).$$
For an exponentially graded material in the $y$-direction, i.e. $\mu=\mu(y)=\mu_0\,e^{2ay}$, it follows from the above relations and definitions that the governing equation for the gradient displacement $u_z$ (recall the Ru-Aifantis theorem [10,11]) reads
\begin{equation}
\left[1-c_1^2\left(\nabla^2+2a\frac{\partial}{\partial
y}\right)\right]\left[1-c_2^2\left(\nabla^2+2a\frac{\partial}{\partial y}\right)\right]u_z=
u_z^0,
\end{equation}
where $c_1^2+c_2^2=\ell^2$, $c_1^2c_2^2=\ell'^4$, and $u_z^0$ denotes the classical elasticity solution discussed in the previous section [Eq.~(9)].

As before, by the substitution $u_z=w\,e^{-ay}$, we obtain
\begin{equation}
\left[1-c_1^2\left(\nabla^2-a^2\right)\right]\left[1-c_2^2\left(\nabla^2-a^2\right)\right]w=w^0
\end{equation}
where $w^0$ is given in Section 2. Use of the two dimensional Fourier transform yields the algebraic equation
\begin{equation}
\left[1+c_1^2\left(s^2+t^2+a^2\right)\right]\left[1+c_2^2\left(s^2+t^2+a^2\right)\right]\tilde{\widetilde{w}}=\tilde{\widetilde{w}}^0,
\end{equation}
where
$$
\notag \tilde{\widetilde{w}}=\mathfrak{F}\{\mathfrak{F}\{w;x\rightarrow
s\};y\rightarrow t\}=\int_{-\infty}^\infty\int_{-\infty}^\infty
w(x,y)\,e^{-\textrm{i}(sx+ty)}ds dt,
$$
and
$$
\tilde{\widetilde{w}}^0=\mathfrak{F}\{\tilde{w}^0;y\rightarrow
t\}=\frac{-\textrm{i}a}{s\,(\omega^2+a^2)}+\frac{t}{s\,(\omega^2+a^2)}-\frac{\pi\delta(s)}{a-\textrm{i}
t};\quad \omega^2=s^2+t^2.
$$
Then the inverse Fourier transform gives
\begin{align}
w=&\frac{1}{(2\pi)^2}\int_{-\infty}^{\infty}\left\{\left(\frac{-\textrm{i}a}{s}+\frac{t}{s}\right)\frac{e^{\textrm{i}(sx+t
y)}}{
\omega^2+a^2}\right\}dsdt\\
\notag
&+\frac{1}{(2\pi)^2}\int_{-\infty}^{\infty}\left\{-\frac{c_1^2}{c_1^2-c_2^2}\,\frac{-\textrm{i}a}{s\left(\omega^2+\kappa_1^2\right)}+\frac{c_2^2}{c_1^2-c_2^2}\,\frac{-\textrm{i}a}{s\left(\omega^2+\kappa^2\right)}\right\}e^{\textrm{i}(sx+t
y)}dsdt\\
\notag
&+\frac{1}{(2\pi)^2}\int_{-\infty}^{\infty}\left\{-\frac{c_1^2}{c_1^2-c_2^2}\,\frac{t}{s\left(\omega^2+\kappa_1^2\right)}+\frac{c_2^2}{c_1^2-c_2^2}\,\frac{t}{s\left(\omega^2+\kappa_2^2\right)}\right\}e^{\textrm{i}
(sx+t
y)}dsdt\\
\notag
&-\frac{\pi}{(2\pi)^2}\int_{-\infty}^{\infty}\left\{\frac{\textrm{i}}{\textrm{i}a+t}-\frac{c_1^2}{c_1^2-c_2^2}\,\frac{a+\textrm{i}t}{t^2+\kappa_1^2}+\frac{c_2^2}{c_1^2-c_2^2}\,\frac{a+\textrm{i}t}{t^2+\kappa_2^2}\right\}e^{\textrm{i}t
y}dt,
\end{align}
where $\kappa_j=\sqrt{a^2+1/c_j^2}$. If we integrate out the variable $s$, and  use the integral relations
$$
\frac{1}{2\pi}\int_{-\infty}^{\infty} \frac{e^{\textrm{i}sx}}{s(s^2+k^2)} ds=\frac{\textrm{i}}{2k^2}\left(1-e^{-|kx|}\right)\sign(x),\quad
\int_0^\infty \frac{t \sin{(t
y)}}{t^2+k^2}\,dt=\frac{\pi}{2}\,\sign(y)\,e^{-|ky|},$$
$$
\int_0^\infty \frac{\cos{(t
y)}}{t^2+k^2}\,dt=\frac{\pi}{2|k|}\,e^{-|ky|},
$$
as well as the symmetry properties of these integrals,  $u_z$ can finally be
expressed in terms of sine and cosine integrals as follows
\begin{align}
u_z=u_z^0&-\frac{c_1^2}{c_1^2-c_2^2}\,\frac{b_z\,e^{-ay}}{2\pi}\,\int_0^\infty
\frac{t\,\sin{(t
y)}}{t^2+\kappa_1^2}\left[\sign(x)\,e^{-\sqrt{t^2+\kappa_1^2}|x|}+2H(-x)\right]dt\\
\notag &+\frac{c_2^2}{c_1^2-c_2^2}\,\frac{b_z\,e^{-ay}}{2\pi}\,\int_0^\infty
\frac{t\,\sin{(ty)}}{t^2+\kappa_2^2}\left[\sign(x)\,e^{-\sqrt{t^2+\kappa_2^2}|x|}+2H(-x)\right]dt\\
\notag &+\frac{c_1^2}{c_1^2-c_2^2}\,\frac{b_z\,e^{-ay}}{2\pi}\,\int_0^\infty
\frac{a\,\cos{(ty)}}{t^2+\kappa_1^2}\left[\sign(x)\,e^{-\sqrt{t^2+\kappa_1^2}|x|}+2H(-x)\right]dt\\
\notag &-\frac{c_2^2}{c_1^2-c_2^2}\,\frac{b_z\,e^{-ay}}{2\pi}\,\int_0^\infty
\frac{a\,\cos{(ty)}}{t^2+\kappa_2^2}\left[\sign(x)\,e^{-\sqrt{t^2+\kappa_2^2}|x|}+2H(-x)\right]dt,
\end{align}
where $u_z^0$ is the classical solution given by Eq.~(9).

It is easily seen that this expression for $a\rightarrow0$ coincides with earlier results obtained by the third author and co-workers for homogeneous media and amended in [12]. It also turns out that for  $\ell=0$ and $\ell'=0$ the result reduces to Eq.(9). When $x\rightarrow 0$, the displacement field can be expressed in an explicit form
\begin{align}
\notag u_z(0,y)=\frac{b_z}{2} H(-y)+\frac{b_z}{4}\,\frac{e^{-ay}}{c_1^2-c_2^2}&\Bigg[-c_1^2\,\sign(y)\,e^{-\kappa_1|y|}+c_2^2\,\sign(y)\,e^{-\kappa_2|y|}\\
&+c_1^2\,\frac{a}{\kappa_1}\,e^{-\kappa_1|y|}-c_2^2\,\frac{a}{\kappa_2}\,e^{-\kappa_2|y|}\Bigg].
\end{align}
It is worth noting that the classical displacement $u_z^0(0,y)$ has an abrupt jump at the dislocation line $y=0$, while the gradient solution of Eq.~(20), is smooth there.
For a fixed value of $a$, the larger the ratio $c_2/c_1$ is, the smoother the solution becomes. It also turns out that the  strains are given by
\begin{align}
\notag \varepsilon_{zx}^{T}=\varepsilon_{zx}^0+\frac{b_z}{4\pi}\,\frac{e^{-ay}}{c_1^2-c_2^2}\Big[-c_1^2\,a\,K_0\left(\kappa_1
r\right)+c_2^2\,a\,K_0\left(\kappa_2
r\right)+c_1^2\,\frac{\kappa_1 y}{r}\,K_1\left(\kappa_1
r\right)\\
-c_2^2\,\frac{\kappa_2 y}{r}\,K_1\left(\kappa_2
r\right)\Big],
\end{align}
\begin{align}
\notag \varepsilon_{zy}^{T}=&\varepsilon_{zy}^0+\frac{b_z}{4\pi}\,\frac{e^{-ay}}{c_1^2-c_2^2}\,\frac{x}{r}\left[-c_1^2
\kappa_1 K_1(\kappa_1 r)+c_2^2 \kappa_2 K_1(\kappa_2 r)\right]\\
\notag
&+\frac{b_z}{4\pi}\,\frac{e^{-ay}\ c_1^2}{c_1^2-c_2^2}\int_0^\infty
\frac{\cos (t
y)}{1+c_1^2\left(t^2+a^2\right)}\,\left[\sign(x)\,e^{-|x|\sqrt{t^2+\kappa_1^2}}+2
H(-x)\right] dt\\
&-\frac{b_z}{4\pi}\,\frac{e^{-ay}\ c_2^2}{c_1^2-c_2^2}\int_0^\infty
\frac{\cos (t
y)}{1+c_2^2\left(t^2+a^2\right)}\,\left[\sign(x)\,e^{-|x|\sqrt{t^2+\kappa_2^2}}+2
H(-x)\right] dt.
\end{align}
where $\varepsilon_{zx}^0$ and $\varepsilon_{zy}^0$ are the classical (elastic) strains given in Section 2. It is seen that $\varepsilon_{zx}^T$ does not contain a plastic part, while $\varepsilon_{zy}^T$ is decomposed into the elastic and plastic strains, i.e. $\varepsilon_{zx}=\varepsilon_{zx}^T$; $\varepsilon_{zx}^P=0,$ and
$$
\varepsilon_{zy}=\varepsilon_{zy}^0+\frac{b_z}{4\pi}\,\frac{e^{-ay}}{c_1^2-c_2^2}\,\frac{x}{r}\left[-c_1^2
\kappa_1 K_1(\kappa_1 r)+c_2^2 \kappa_2 K_1(\kappa_2 r)\right],\quad
\varepsilon_{zy}^P=\varepsilon_{zy}^T-\varepsilon_{zy}.
$$
It is also follows that the expression for the plastic strain $\varepsilon_{zy}^P(0,y)$ is given by the simple formula
$$
\varepsilon_{zy}^P(0,y)=\frac{b_z}{8}\,\frac{e^{-ay}}{c_1^2-c_2^2}\left[\frac{e^{-\kappa_1|y|}}{\kappa_1}-\frac{e^{-\kappa_2|y|}}{\kappa_2}\right].
$$

The lower-order elastic-like stresses and higher-order double stresses are given by the expressions:
\begin{align}
\notag \sigma_{zx}^E&=\sigma_{zx}^0+\frac{b_z\,\mu_0\,e^{ay}}{2\pi\left(c_1^2-c_2^2\right)}\Big[-c_1^2\,a\,K_0\left(\kappa_1
r\right)+c_2^2\,a\,K_0\left(\kappa_2
r\right)+c_1^2\,\frac{\kappa_1 y}{r}\,K_1\left(\kappa_1
r\right)-c_2^2\,\frac{\kappa_2 y}{r}\,K_1\left(\kappa_2
r\right)\Big]\\
\sigma_{zy}^E&=\sigma_{zy}^0+\frac{b_z\,\mu_0\,e^{ay}}{2\pi\left(c_1^2-c_2^2\right)}\,\frac{x}{r}\left[-c_1^2
\kappa_1 K_1(\kappa_1 r)+c_2^2 \kappa_2 K_1(\kappa_2 r)\right],
\end{align}
where $\sigma_{zx}^0$ and $\sigma_{zy}^0$ are the classical stresses given by Eqs.~(12), and
\begin{align}
\notag \tau_{(zx)x}=&2\ell^2\,\mu_0\,e^{2ay}\varepsilon_{zx,x}\\
\notag
=&-a\ell^2\sigma_{zx}^E+\frac{b_z\,\ell^2\,\mu_0}{2\pi}\,e^{ay}\Bigg\{\frac{b^2xy}{r^2}
K_0(ar)+\left(\frac{2axy}{r^3}-\frac{a^2
x}{r}\right)K_1(ar)\\
\notag &-\frac{1}{c_1^2-c_2^2}\Bigg[\frac{xy
}{r^2}\left[c_1^2\kappa_1^2\,K_0(\kappa_1 r)-c_2^2 \kappa_2^2\,
K_0(\kappa_2 r)\right]\\
\notag &\hspace{2cm} +\left(\frac{2 xy}{r^3}-\frac{a
x}{r}\right)\left[c_1^2 \kappa_1\,K_1(\kappa_1 r) -c_2^2
\kappa_2K_1(\kappa_2 r)\right]\Bigg]\Bigg\},\\
\notag\tau_{(zx)y}=&2\ell^2\,\mu_0\,e^{2ay}\varepsilon_{zx,x}\\
\notag
=&-a\ell^2\sigma_{zx}^E+\frac{b_z\ell^2\,\mu_0}{2\pi}\,e^{ay}\Bigg\{\frac{a^2y^2}{r^2}
K_0(ar)+\left(\frac{2ay^2}{r^3}-\frac{a^2
y}{r}-\frac{a}{r}\right)K_1(ar)\\
\notag &-\frac{1}{c_1^2-c_2^2}\Bigg[\frac{y^2
}{r^2}\left[c_1^2\kappa_1^2\,K_0(\kappa_1 r)-c_2^2 \kappa_2^2\,
K_0(\kappa_2 r)\right]\\
&\hspace{2cm} +\left(\frac{2 y^2}{r^3}-\frac{a
y}{r}-\frac{1}{r}\right)\left[c_1^2 \kappa_1\,K_1(\kappa_1 r)
-c_2^2 \kappa_2K_1(\kappa_2 r)\right]\Bigg]\Bigg\},\\
\notag \tau_{(zy)x}=&2\ell^2\,\mu_0\,e^{2ay}\varepsilon_{zy,x}\\
\notag
=&-a\ell^2\sigma_{zy}^E-\frac{b_z\,\ell^2\,\mu_0}{2\pi}\,e^{ay}\Bigg\{\frac{a^2x^2}{r^2}
K_0(br)+\left(\frac{2ax^2}{r^3}-\frac{a}{r}\right)K_1(ar)\\
\notag &-\frac{1}{c_1^2-c_2^2}\Bigg[\frac{x^2
}{r^2}\left[c_1^2\kappa_1^2\,K_0(\kappa_1 r)-c_2^2 \kappa_2^2\,
K_0(\kappa_2 r)\right]\\
\notag &\hspace{2cm} +\left(\frac{2
x^2}{r^3}-\frac{1}{r}\right)\left[c_1^2 \kappa_1\,K_1(\kappa_1 r)
-c_2^2 \kappa_2K_1(\kappa_2 r)\right]\Bigg]\Bigg\},\\
\notag \tau_{(zy)y}=&2\ell^2\,\mu_0\,e^{2ay}\varepsilon_{zy,y}\\
\notag
&=-a\ell^2\sigma_{zy}^E-\frac{b_z\,\ell^2\,\mu_0}{2\pi}\,e^{ay}\Bigg\{\frac{a^2xy}{r^2}
K_0(ar)+\frac{2axy}{r^3}\,K_1(ar)\\
\notag &-\frac{1}{c_1^2-c_2^2}\Bigg[\frac{xy
}{r^2}\left[c_1^2\kappa_1^2\,K_0(\kappa_1 r)-c_2^2 \kappa_2^2\,
K_0(\kappa_2 r)\right]+\frac{2 xy}{r^3}\,\left[c_1^2
\kappa_1\,K_1(\kappa_1 r) -c_2^2 \kappa_2K_1(\kappa_2
r)\right]\Bigg]\Bigg\}.
\end{align}
It is seen from the above expressions that $\sigma_{zx}^E$ is still symmetric with respect to the plane $x=0$, while $\sigma_{zy}^E$ has lost symmetry with respect to plane $y=0$. Moreover, in contrast to homogeneous medium, $\tau_{(zx)x}\neq-\tau_{(zy)y}$. More details and for this problem their physical implications to possible improvements of designing FGMs and the expressions for the triple stresses will be given in a forthcoming publication.

\par\bigskip\noindent
\textbf{References}

\noindent [1] Lazar, M., 2007. On a screw dislocation in a functionally graded material, Mech. Res. Comm. 34, 305-311.\\
\noindent [2] Kioseoglou, J., Dimitrakopulos, G.P., Komninou, Ph., Karakostas, Th., Aifantis, E.C., 2008. Dislocation core investigation by geometric phase analysis and the dislocation density tensor, J. Phys. D: Appl. Phys. 41, 035408.\\
\noindent [3] Kioseoglou, J., Konstantopoulos, I., Ribarik, G., Dimitrakopulos, G.P., Aifantis, E.C., 2009. Nonsingular dislocation and crack fields: implications to small volumes, Microsyst. Techol. 15, 117-121.\\
\noindent [4] Chan, Y., Paulino, G.H., Fannjiang, A.C., 2008. Gradient elasticity theory for mode III fracture in functionally graded materials---part II: Crack parallel to the material gradation. J. Appl. Mech. 75, 061015.\\
\noindent [5] Paulino, G.H., Fannjiang, A.C., Chan, Y.-S., 2003. Gradient elasticity theory for mode III fracture in functionally graded materials---part I: Crack perpendicular to the material gradation. J. of Appl. Mech. 70, 531-542.\\
\noindent [6] Erdogan, F., 1995. Fracture mechanics of functionally graded materials. Compos. Eng. 5, 753-770.\\
\noindent [7] Markworth, A.J., Ramesh, K.S., Parks, W.P., 1995. Modeling studies applied to functionally graded materials, J. Mater. Sci. 30, 2183-2193.\\
\noindent [8] Suresh, S., Mortensen, A., 1998. Fundamentals of functionally graded materials, ASM International and the Institute of Materials, IOM Communications Ltd., London.\\
\noindent[9] Lazar, M., Maugin, G.A., Aifantis, E.C., 2006. Dislocations in second strain gradient elasticity, Int. J. Solids Struct. 43, 1787-1817.\\
\noindent [10] Ru, C.Q., Aifantis, E.C., 1993. A simple approach to solve boundary value problems in gradient elasticity, Acta Mech. 101, 59-68.\\
\noindent [11] Aifantis, E.C., 2003. Update on a class of gradient theories, Mech. Mater. 35, 259-280.\\
\noindent [12] Lazar, M., Maugin, G.A., 2006. Dislocations in gradient elasticity revisited. Proc. R. Soc. A 462, 3465-3480.

\end{document}